\documentclass[aps,english,prl,floatfix,amsmath,superscriptaddress,tightenlines,twocolumn,nofootinbib]{revtex4-2}
\usepackage{mathtools, amssymb}
\usepackage{graphicx}
\usepackage{tikz}
\usepackage{tikz-3dplot}
\usetikzlibrary{spy}
\usetikzlibrary{arrows.meta}
\usetikzlibrary{calc}
\usepackage{intcalc}
\usepackage[caption=false]{subfig}
\usepackage[shortlabels]{enumitem}
\usepackage{breakurl}

\usepackage{multirow}
\usepackage{hhline}
\usepackage{algorithmicx}
\usepackage{algorithm}
\usepackage[noend]{algpseudocode}

\usepackage{soul}
\usepackage[utf8]{inputenc}
\usepackage{xcolor}
\usepackage{amsthm}
\usepackage{float}
\usepackage{comment}
\usepackage{qcircuit}

\newcommand{\Tr}[0]{{\text{Tr}}}
\newcommand{\fTr}[0]{\widetilde{\text{Tr}}}

\newcommand{\coloredboxhx}[5]{
	\foreach \x in {#2, ..., #4}
	{
		\foreach \y in {#3, ..., #5}
		{
			\fill[#1] (\x-0.5*\y +0.5, \y*0.866 -0.289) -- (\x-0.5*\y +0.5, \y*0.866 +0.289) -- (\x-0.5*\y, \y*0.866+0.577) -- (\x-0.5*\y -0.5, \y*0.866 +0.289) --(\x-0.5*\y-0.5, \y*0.866 -0.289)--(\x-0.5*\y, \y*0.866 -0.577)--(\x-0.5*\y+0.5, \y*0.866 -0.289) -- cycle;
			
			\draw (\x-0.5*\y +0.5, \y*0.866 -0.289) -- (\x-0.5*\y +0.5, \y*0.866 +0.289) -- (\x-0.5*\y, \y*0.866+0.577) -- (\x-0.5*\y -0.5, \y*0.866 +0.289) --(\x-0.5*\y-0.5, \y*0.866 -0.289)--(\x-0.5*\y, \y*0.866 -0.577)--(\x-0.5*\y+0.5, \y*0.866 -0.289) -- cycle;
		}
	}
}

\newcommand{\coloredkv}[4]{
	\begin{scope}[xshift=#3 cm, xshift=-0.5*#4 cm, yshift=0.866*#4 cm, rotate=60*#2]
		\fill[#1] (-30:0.578) -- (30:0.578) -- (90:0.578) -- (150:0.578) --(210:0.578)--(270:0.578) -- cycle;
		
		\draw (-30:0.578) -- (30:0.578) -- (90:0.578) -- (150:0.578) --(210:0.578)--(270:0.578) -- cycle;
	\end{scope}	
}
\newcommand{\coloredke}[4]{
	\begin{scope}[xshift=#3 cm, xshift=-0.5*#4 cm, yshift=0.866*#4 cm, rotate=60*#2]
		\fill[#1] (-30:0.578) -- (30:0.578) -- (90:0.578) -- (150:0.578) --(210:0.578)--(270:0.578) -- cycle;
		\fill[#1,xshift=1 cm] (-30:0.578) -- (30:0.578) -- (90:0.578) -- (150:0.578) --(210:0.578)--(270:0.578) -- cycle;
		
		\draw (30:0.578) -- (90:0.578) -- (150:0.578) --(210:0.578)--(270:0.578)--(330:0.578)--++(-30:0.578)--++(30:0.578)--++(90:0.578)--++(150:0.578)--++(210:0.578)--cycle;
	\end{scope}	
}
\newcommand{\coloredkf}[4]{
	\begin{scope}[xshift=#3 cm, xshift=-0.5*#4 cm, yshift=0.866*#4 cm, rotate=60*#2]
		\fill[#1] (-30:0.578) -- (30:0.578) -- (90:0.578) -- (150:0.578) --(210:0.578)--(270:0.578) -- cycle;
		\fill[#1,xshift=1 cm] (-30:0.578) -- (30:0.578) -- (90:0.578) -- (150:0.578) --(210:0.578)--(270:0.578) -- cycle;
		\fill[#1,xshift=0.5 cm,yshift=0.866 cm] (-30:0.578) -- (30:0.578) -- (90:0.578) -- (150:0.578) --(210:0.578)--(270:0.578) -- cycle;
		
		\draw (90:0.578) -- (150:0.578) --(210:0.578)--(270:0.578)--(330:0.578)--++(-30:0.578)--++(30:0.578)--++(90:0.578)--++(150:0.578)
		--++(90:0.578)--++(150:0.578)--++(210:0.578)--++(270:0.578)--cycle;
	\end{scope}	
}

\newcommand{\diskshadded}[6]
{
	\foreach \x in {#1,...,#3}
	{
		\foreach \y in {#2,...,#4}
		{
			
			\fill[#6] (\x-0.5*\y +0.5, \y*0.866 -0.289) -- (\x-0.5*\y +0.5, \y*0.866 +0.289) -- (\x-0.5*\y, \y*0.866+0.577) -- (\x-0.5*\y -0.5, \y*0.866 +0.289) --(\x-0.5*\y-0.5, \y*0.866 -0.289)--(\x-0.5*\y, \y*0.866 -0.577)--(\x-0.5*\y+0.5, \y*0.866 -0.289) -- cycle;
			
			\draw[color=#5] (\x-0.5*\y +0.5, \y*0.866 -0.289) -- (\x-0.5*\y +0.5, \y*0.866 +0.289) -- (\x-0.5*\y, \y*0.866+0.577) -- (\x-0.5*\y -0.5, \y*0.866 +0.289) --(\x-0.5*\y-0.5, \y*0.866 -0.289)--(\x-0.5*\y, \y*0.866 -0.577)--(\x-0.5*\y+0.5, \y*0.866 -0.289) -- cycle;
		}
	}
}
\newcommand{\wordbox}[4]{	
		\foreach \x in {#2, ..., #2}
	{
		\foreach \y in {#3, ..., #3}
		{
		\node at (\x-0.5*\y, 0.866*\y) {\color{#1}{#4}};
					}
		}
}

\definecolor{maroon}{rgb}{144,12,63}
\definecolor{darkblue}{rgb}{27,12,144}
\definecolor{mypurple2}{RGB}{170,0,255}
\definecolor{myred}{RGB}{255, 0, 90}
\definecolor{mycyan}{RGB}{0, 191, 255}
\definecolor{KK}{RGB}{0, 180, 20}

\definecolor{BSorange}{RGB}{140,50,0}

\usepackage{hyperref}
\hypersetup{colorlinks=true,linkcolor=myred,citecolor=magenta,urlcolor=blue}
\usepackage[all]{hypcap}

\begin{document}
\author{Isaac H. Kim}
\thanks{Equal contribution.}
\affiliation{Department of Computer Science, University of California, Davis, Davis, CA, USA}
\affiliation{Centre for Engineered Quantum Systems, School of Physics, University of Sydney, Sydney, NSW, Australia}
\author{Bowen Shi}
\thanks{Equal contribution.}
\affiliation{Department of Physics, University of California at San Diego, La Jolla, CA, USA}
\author{Kohtaro Kato}
\affiliation{Center for Quantum Information and Quantum Biology, Osaka University, Osaka, Japan}
\affiliation{Graduate School of Informatics, Nagoya University, Nagoya, Japan}
\author{Victor V. Albert}
\affiliation{Joint Center for Quantum Information and Computer Science, National Institute of Standards and Technology and University of Maryland, College Park, MD, USA}

\title{Chiral central charge from a single bulk wave function}
\date{\today}

\begin{abstract} 
A $(2+1)$-dimensional gapped quantum many-body system can have a topologically protected energy current at its edge. The magnitude of this current is determined entirely by the temperature and the chiral central charge, a quantity associated with the effective field theory of the edge. We derive a formula for the chiral central charge that, akin to the topological entanglement entropy, is completely determined by the many-body ground state wave function in the bulk. According to our formula, nonzero chiral central charge gives rise to a topological obstruction that prevents the ground state wave function from being real-valued in any local product basis.
\end{abstract}
\maketitle

Topological order \cite{WEN1990a,Kitaev2003} is an active topic of investigation today, simultaneously admitting connections to deep mathematics and the promise of intrinsically protected quantum devices. Well-known properties of topologically ordered systems in $(2+1)$ dimensions are anyonic excitations in the bulk~\cite{Leinaas1977,Wilczek1982,Arovas1984} and robust gapless modes on the edge of the system~\cite{Thouless1982,Wen1991,Kane1997}, both observed in fractional quantum Hall systems experimentally~\cite{Nakamura2020,Klitzing1980,FQHE}. 

A remarkable observation is that the features characterizing the low-energy excitations of a system can be extracted from its many-body ground-state wave function. Early examples of this observation include topological entanglement entropy~\cite{Kitaev2006,Levin2006} and the entanglement spectrum~\cite{Li2008}. Later work has gone on to extract other related properties, \emph{e.g.,} topological $S$- and $T$-matrices of the anyonic excitations~\cite{Zhang2012,Cincio2013}. 

These properties, however, do not exhaustively constitute the data that defines the phase of the underlying system. A missing piece of information is the \emph{chiral central charge} (CCC), denoted as $c_-$ here. A nonzero $c_-$ implies the presence of gapless edge excitations, so a natural place in which the CCC appears is the physical edge of the system. At a temperature $T$ that is low compared to the bulk excitation gap, the energy current $I$ along the edge is~\cite{Kane1997,Senthil1999,Read2000,Cappelli2002,Kitaev2006a} 
\begin{equation}
    I=\frac{\pi}{12}c_- T^2. \label{eq:edge_current}
\end{equation}

A prominent physical system with nonzero CCC is the two-dimensional electron gas in a magnetic field~\cite{Klitzing1980,Thouless1982,FQHE}. Besides admitting the well-known quantized electrical response protected by $U(1)$ charge conservation symmetry, this system admits a quantized \textit{thermal} response~\cite{Kane1997} as well, which manifests as a unidirectional edge energy current. This current corresponds to a nonzero $c_-$ and exists independently of the $U(1)$ symmetry~\cite{Read2000,Senthil1999,Cappelli2002}. More generally, the chiral central charge is an integer when the edge admits a chiral Luttinger liquid description~\cite{Wen1991}, but can be a rational number in, for instance, $p+ip$ superconductors and systems with non-Abelian anyons~\cite{Read2000,Moore1991}.

While the CCC describes a physical property of the edge, it is also related to certain properties of the bulk. For instance, the energy current at the edge can be related to a $2$-current in the bulk, which can be computed from a microscopic Hamiltonian~\cite{Kitaev2006a,Kapustin2020}. In effective field theory approaches, CCC  appears in the gravitational Chern-Simons term of the bulk action, which is responsible for the framing anomaly of the underlying system~\cite{Gromov2015}. This lets us relate CCC to the topological Berry phase under adiabatic variation of the metric~\cite{Bradlyn2015} and the Hall viscosity on a sphere~\cite{Abanov2014,Klevtsov2015,Bradlyn2015,Golan2019}. Moreover, given a set of ground state wave functions on a torus, CCC can be computed (up to a fixed integer) using topological $S$- and $T$-matrices~\cite{Zhang2012} or momentum polarization~\cite{Tu2013}. Alternatively, CCC can be inferred from the entanglement spectrum of the bulk reduced density matrix on a disk~\cite{Li2008}. However, a succinct closed-form formula that relates CCC to the ground state entanglement of a single wave function --- akin to the topological entanglement entropy~\cite{Kitaev2006,Levin2006} --- has been missing. 

In this Letter, we introduce a new formula that reveals a connection between CCC and the entanglement structure of the bulk. Our formula is based on the modular commutator  --- a new quantity expressed in terms of reduced density matrices of a many-body wave function. We argue that the CCC can be expressed in terms of the modular commutator for ground states of gapped Hamiltonians,
with or without symmetries. We numerically confirm our formula up to a small error attributable to finite-size effects in our companion paper~\cite{Long}.

\emph{Summary of results---}For a general tripartite state $\rho_{ABC}$ on a finite dimensional Hilbert space, the \textit{modular commutator} is
\begin{equation}
    J(A,B,C)_{\rho} := i\text{Tr}\left(\rho_{ABC}[K_{AB}, K_{BC}] \right),\label{eq:J}
\end{equation}
where $K_A = -\ln \rho_A$ is the modular Hamiltonian~\cite{Haag2012,Casini2008} associated with the reduced density matrix $\rho_A$ on subsystem $A$. We can readily see that $J$ is real (because $[K_{AB}, K_{BC}]$ is an antihermitian operator) and odd under complex conjugation (with respect to any product-state basis over the local degrees of freedom). The latter operation corresponds to time reversal in our physical context, meaning that it can flip the direction of a system's edge current. Thus, the fact that $J$ is odd under time reversal is a necessary property for it to encode information about the CCC.

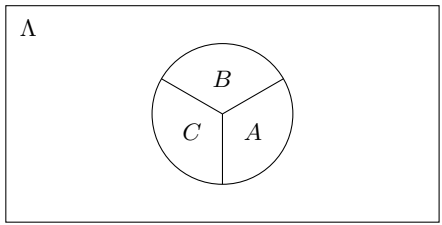
\begin{figure}[h]
	\centering
	\begin{tikzpicture}[scale=0.36]
		\draw[] (-8, -4) -- (8,-4) -- (8,4) -- (-8,4) -- cycle;
		\node[below right] (lambda) at (-7.8,3.8) {$\Lambda$};
		\draw (0,0) circle (2.6);
		\draw (0,0) -- (30:2.6);
		\draw (0,0) -- (150:2.6);
		\draw (0,0) -- (-90:2.6);
		\node at (90:1.3) {$B$};
		\node at (-150:1.3) {$C$};
		\node at (-30:1.3) {$A$};
	\end{tikzpicture}
	\caption{Partition of a disk-shaped region $ABC$ in the bulk ($\Lambda$). Each subsystem is assumed to be sufficiently large compared to the correlation length.}
	\label{fig:abc}
\end{figure}

Plugging in a many-body ground state $\sigma = |\psi\rangle\!\langle \psi|$ satisfying the area law with a constant subcorrection term~\cite{Kitaev2006,Levin2006}, we relate the modular commutator \eqref{eq:J} to the edge energy current \eqref{eq:edge_current}, obtaining our main result
\begin{equation}
J(A,B,C)_\sigma= \frac{\pi}{3} c_- \label{eq:main_result}
\end{equation}
for subsystems $A$, $B$, and $C$ depicted in Fig.~\ref{fig:abc}. 
Eq.~\eqref{eq:main_result} is insensitive to continuous deformations of the subsystems, so long as they remain to partition a disk. 

In the rest of this Letter, we derive Eq.~\eqref{eq:main_result} using theoretical tools developed in Refs.~\cite{SKK2019,Shi2021,Shi2021a}. Using the fact that $\sigma$ satisfies the area law, we show that the state's modular Hamiltonian $K_\mathfrak{D}$ for a disk $\mathfrak{D}$ can be decomposed into a sum of local operators. From this decomposition, we can obtain an expression for the ``energy current''. The value of the ``energy current'' is determined by the expectation values of commutators of the local terms of $K_\mathfrak{D}$, and each nonzero contribution is shown to be of the form of the modular commutator~\eqref{eq:J}, up to a proportionality constant. By viewing $K_\mathfrak{D}$ as a physical Hamiltonian describing the same phase as that of $\sigma$, we relate $K_\mathfrak{D}$'s ``energy current'' to the physical current from Eq.~\eqref{eq:edge_current}, yielding Eq.~\eqref{eq:main_result}.

\emph{Markov-chain states---}An important fact about the modular commutator is that it vanishes if the underlying state is a quantum Markov chain~\cite{Petz2003}. Specifically, consider a tripartite state $\rho_{XYZ}$. This state is a quantum Markov chain if its conditional mutual information --- defined as $I(X:Y|Z)_{\rho} := S(\rho_{XY}) + S(\rho_{YZ}) - S(\rho_Y)-S(\rho_{XYZ})$, where $S(\rho):= -\text{Tr}(\rho \ln \rho)$ is the von Neumann entropy of $\rho$ --- is zero. It turns out that
\begin{equation}
    I(X:Z|Y)_{\rho}=0 \implies J(X,Y,Z)_{\rho}=0~,
\label{eq:cmi_Jabc}
\end{equation}
which can be proved via the important relation~\cite{Petz2003}\footnote{This is true for positive definite $\rho_{XYZ}$. If $\rho_{XYZ}$ has zero eigenvalues, Eq.~(\ref{eq:markov_decomposition}) should be replaced by $ K_{XYZ}\,\rho_{XYZ} = (K_{XY} + K_{YZ} - K_Y)\rho_{XYZ}$, meaning that the same condition holds on the subspace spanned by the eigenstates of the nonzero eigenvalues of $\rho_{XYZ}$; see the Supplemental Material for details, in particular, on the derivation of Eq.~\eqref{eq:markov_decomposition} for fermions.} 
\begin{equation}
    I(X:Z|Y)_{\rho}=0\Longleftrightarrow K_{XYZ}=K_{XY}+K_{YZ}-K_{Y}\label{eq:markov_decomposition}.
\end{equation}
Applying this relation, the modular commutator becomes $J(X,Y,Z) = i\text{Tr}(\rho_{XYZ}[K_{XY}, K_{XYZ} +K_Y])$. The cyclicity of the trace then implies that this expression is zero. 

\emph{Area law and modular commutator---}Now we shift our focus to ground states of gapped quantum many-body systems in two spatial dimensions. Such states, which we denote as $\sigma$, are expected to obey the area law of entanglement entropy~\cite{Kitaev2006,Levin2006}, which means that the following equation holds for any disk-shaped region $A$:
\begin{equation}
    S(\sigma_A) = \alpha |\partial A| - \gamma + \ldots, \label{eq:area_law}
\end{equation}
where $|\partial A|$ is the length of the perimeter of $A$, $\alpha$ is a non-universal constant, and $\gamma$ is the topological entanglement entropy~\cite{Kitaev2006,Levin2006}. The remaining term vanishes in the $|A|\to \infty$ limit.

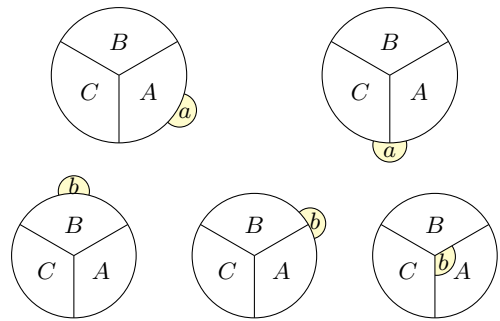
\begin{figure}[h]
	\centering 
	\begin{tikzpicture}
		\begin{scope}[scale=0.3]
			% 1st row: primary cases
			\begin{scope}
				\fill[yellow, opacity=0.2] (-30:3.1) circle (0.75);
				\draw(-30:3.1) circle (0.75);
				
				\draw[fill=white] (0,0) circle (3);
				\node at (90:1.5) {$B$};
				\node at (-30:1.5) {$A$};
				\node at (-150:1.5) {$C$};
				\draw (0:0)--(30:3);
				\draw (0:0)--(150:3);
				\draw (0:0)--(-90:3);
				\node (a) at (-30:3.375) {$a$};
			\end{scope}	
			\begin{scope}[xshift=12 cm]
				\fill[yellow, opacity=0.2] (-90:3.1) circle (0.75);
				\draw(-90:3.1) circle (0.75);
				
				\draw[fill=white] (0,0) circle (3);
				\node at (90:1.5) {$B$};
				\node at (-30:1.5) {$A$};
				\node at (-150:1.5) {$C$};
				\draw (0:0)--(30:3);
				\draw (0:0)--(150:3);
				\draw (0:0)--(-90:3);
				\node (a) at (-90:3.378) {$a$};
			\end{scope}	
		\end{scope}	
		%2nd row: cases that can be reduced to those in the 1st row
		\begin{scope}[yshift=-2.4 cm, scale=0.30]
			\begin{scope}[xshift=-2 cm, scale=0.92]
				\fill[yellow, opacity=0.2] (90:3.1) circle (0.75);
				\draw(90:3.1) circle (0.75);
				
				\draw[fill=white] (0,0) circle (3);
				\node at (90:1.5) {$B$};
				\node at (-30:1.5) {$A$};
				\node at (-150:1.5) {$C$};
				\draw (0:0)--(30:3);
				\draw (0:0)--(150:3);
				\draw (0:0)--(-90:3);
				\node (a) at (90:3.375) {$b$};
			\end{scope}	
			\begin{scope}[xshift=6 cm, scale=0.92]
				\fill[yellow,opacity=0.2] (30:3.1) circle (0.75);
				\draw(30:3.1) circle (0.75);
				
				\draw[fill=white] (0,0) circle (3);
				\node at (90:1.5) {$B$};
				\node at (-30:1.5) {$A$};
				\node at (-150:1.5) {$C$};
				\draw (0:0)--(30:3);
				\draw (0:0)--(150:3);
				\draw (0:0)--(-90:3);
				\node (a) at (30:3.375) {$b$};
			\end{scope}	
			\begin{scope}[xshift=14 cm, scale=0.92]
				\fill[yellow,opacity=0.2] (-30:0.1) circle (0.9);
				\draw (-30:0.1) circle (0.9);
				\fill[white] (0:0)--(30:2)--(150:2)--(-90:2)--(0:0)--cycle;
				
				\draw (0,0) circle (3);
				\node at (90:1.5) {$B$};
				\node at (-30:1.5) {$A$};
				\node at (-150:1.5) {$C$};
				\draw (0:0)--(30:3);
				\draw (0:0)--(150:3);
				\draw (0:0)--(-90:3);
				\node (a) at (-34:0.5) {$b$};
			\end{scope}	
		\end{scope}			
	\end{tikzpicture}
	\caption{First row: deformation of $A$ to $Aa$, where $a$ is a small disk separated from $B$. Second row: deformations of $B$ to $Bb$; (for the rightmost figure, $b\subset A$). All cases in the second row can be reduced to the cases in the first row by a change of subsystem.}
	\label{fig:deformation}
\end{figure}

The area law implies that the modular commutator $J(A,B,C)$ for the partition of the disk in Fig.~\ref{fig:abc} is a topological invariant. Specifically, consider the deformations of subsystems $A, B,$ and $C$ as described in Fig.~\ref{fig:deformation}. 
Deformations away from $B$ leaves the modular commutator invariant for the following reason. Without loss of generality, consider $a \subset \Lambda\setminus (ABC)$ as is shown in the first row of Fig.~\ref{fig:deformation}. From  $I(a:B\vert A)=0$ and Eq.~\eqref{eq:markov_decomposition}, we can obtain  $K_{aAB} = K_{AB} + K_{aA} - K_{A}$; here the vanishing of conditional mutual information $I(a:B\vert A)=0$ follows from  Eq.~(\ref{eq:area_law}). Therefore, $J(Aa,B,C) = J(A,B,C)$. By repeating the same argument, one can freely deform the part of the edges and the triple intersection point that is separated from $B$.

The deformations of $B$ also leaves the modular commutator invariant. To establish this fact, it suffices to consider the deformations depicted in the second row of Fig.~\ref{fig:deformation} and show that $J(A,B,C)=J(A\setminus b,Bb,C)$. While the argument in the previous paragraph does not apply directly, it does apply after we switch the subsystems as follows. Let $\Lambda=ABCD$ and suppose $\sigma=\vert \psi\rangle \langle \psi \vert$ is pure. 
Then it follows from $K_{BC}\vert \psi\rangle = K_{AD} \vert \psi\rangle$ that $J(A,B,C)=J(B, A, D)$; similarly, we can put either $C$ or $D$ in the middle entry and have $J(A,B,C)=J(\cdot, C, \cdot)=J(\cdot, D, \cdot)$ with appropriate choices of the first and third entries. Thus, we can always make the middle entry to be away from the place that the deformation occurs. The argument in the previous paragraph now applies. Therefore, given subsystems $A,B,$ and $C$ which are topologically equivalent to Fig.~\ref{fig:abc}, $J(A,B,C)$ is invariant under any smooth deformation of $A,B,$ or $C$. 

\emph{Locality of modular Hamiltonian---}We now discuss the ``local'' structure of the modular Hamiltonian $K_\mathfrak{D}$ of a many-body area-law state $\sigma$
on a disk-shaped region $\mathfrak{D}$. The area law implies that
$I(X:Z|Y)_{\sigma}=0$ for any ``chain-like'' region $XYZ$, \emph{i.e.,} for which $X,Y,Z$
are simply connected, and $X$ and $Z$ do not share a boundary.
Partitioning the disk $\mathfrak{D}=XYZ$ into such a region, Eq.~\eqref{eq:markov_decomposition}
allows us to express the modular Hamiltonian $K_{\mathfrak{D}}$ using
terms of smaller support. This process can then be applied recursively
to the resulting terms $\{K_{XY},K_{YZ},K_{Y}\}$, yielding
an ever-more local decomposition for $K_{\mathfrak{D}}$. 

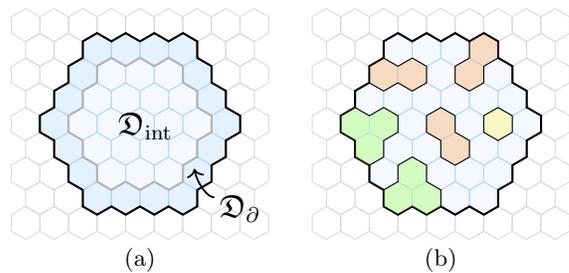
\begin{figure}[h]
	\centering
	\begin{tikzpicture}
		\begin{scope}[scale=0.38]
			% 0th line
			\diskshadded{-2}{0}{6}{0}{gray!30!white}{white,opacity=0.4};
			%1st line
			\diskshadded{-1}{1}{0}{1}{gray!30!white}{white,opacity=0.4};
			\diskshadded{1}{1}{4}{1}{blue!50!cyan!30!white}{blue!50!cyan!10!white,opacity=0.4};
			\diskshadded{5}{1}{6}{1}{gray!30!white}{white,opacity=0.4};
			%2nd line
			\diskshadded{-1}{2}{0}{2}{gray!30!white}{white,opacity=0.4};
			\diskshadded{1}{2}{5}{2}{blue!50!cyan!30!white}{blue!50!cyan!10!white,opacity=0.4};
			\diskshadded{6}{2}{7}{2}{gray!30!white}{white,opacity=0.4};
			%3rd line
			\diskshadded{0}{3}{0}{3}{gray!30!white}{white,opacity=0.4};
			\diskshadded{1}{3}{6}{3}{blue!50!cyan!30!white}{blue!50!cyan!10!white,opacity=0.4};
			\diskshadded{7}{3}{7}{3}{gray!30!white}{white,opacity=0.4};
			%4th line
			\diskshadded{0}{4}{0}{4}{gray!30!white}{white,opacity=0.4};
			\diskshadded{1}{4}{7}{4}{blue!50!cyan!30!white}{blue!50!cyan!10!white,opacity=0.4};
			\diskshadded{8}{4}{8}{4}{gray!30!white}{white,opacity=0.4};
			%5th line
			\diskshadded{1}{5}{1}{5}{gray!30!white}{white,opacity=0.4};
			\diskshadded{2}{5}{7}{5}{blue!50!cyan!30!white}{blue!50!cyan!10!white,opacity=0.4};
			\diskshadded{8}{5}{8}{5}{gray!30!white}{white,opacity=0.4};
			%6th line
			\diskshadded{1}{6}{2}{6}{gray!30!white}{white,opacity=0.4};
			\diskshadded{3}{6}{7}{6}{blue!50!cyan!30!white}{blue!50!cyan!10!white,opacity=0.4};
			\diskshadded{8}{6}{9}{6}{gray!30!white}{white,opacity=0.4};
			%7st line
			\diskshadded{2}{7}{3}{7}{gray!30!white}{white,opacity=0.4};
			\diskshadded{4}{7}{7}{7}{blue!50!cyan!30!white}{blue!50!cyan!10!white,opacity=0.4};
			\diskshadded{8}{7}{9}{7}{gray!30!white}{white,opacity=0.4};
			% 8th line
			\diskshadded{2}{8}{10}{8}{gray!30!white}{white,opacity=0.4};
			
			% darker D_partial
			\diskshadded{1}{1}{4}{1}{blue!50!cyan!30!white}{blue!50!cyan!20!white,opacity=0.4};
			\diskshadded{1}{2}{1}{4}{blue!50!cyan!30!white}{blue!50!cyan!20!white,opacity=0.4};
			\diskshadded{7}{4}{7}{7}{blue!50!cyan!30!white}{blue!50!cyan!20!white,opacity=0.4};
			\diskshadded{4}{7}{6}{7}{blue!50!cyan!30!white}{blue!50!cyan!20!white,opacity=0.4};
			\diskshadded{3}{6}{3}{6}{blue!50!cyan!30!white}{blue!50!cyan!20!white,opacity=0.4};
			\diskshadded{2}{5}{2}{5}{blue!50!cyan!30!white}{blue!50!cyan!20!white,opacity=0.4};
			\diskshadded{5}{2}{5}{2}{blue!50!cyan!30!white}{blue!50!cyan!20!white,opacity=0.4};
			\diskshadded{6}{3}{6}{3}{blue!50!cyan!30!white}{blue!50!cyan!20!white,opacity=0.4};
			
			% draw the outer line
			\draw[line width=0.8 pt] (30:0.578) -- (90:0.578) 
			-- ++(90:0.578)
			-- ++(150:0.578)-- ++(90:0.578) -- ++(150:0.578)-- ++(90:0.578) -- ++(150:0.578)-- ++(90:0.578) 
			-- ++(30:0.578)--++(90:0.578) -- ++(30:0.578)--++(90:0.578) -- ++(30:0.578)--++(90:0.578) -- ++(30:0.578)
			--++(-30:0.578) -- ++(30:0.578)--++(-30:0.578) -- ++(30:0.578)--++(-30:0.578) -- ++(30:0.578)--++(-30:0.578)
			-- ++(-90:0.578)--++(-30:0.578) -- ++(-90:0.578)--++(-30:0.578) -- ++(-90:0.578)--++(-30:0.578) -- ++(-90:0.578)
			--++(-150:0.578)--++(-90:0.578)--++(-150:0.578)--++(-90:0.578)--++(-150:0.578)--++(-90:0.578)--++(-150:0.578)
			--++(150:0.578)--++(210:0.578)--++(150:0.578)--++(210:0.578)--++(150:0.578)--++(210:0.578)--cycle;
			
			% draw the inner line
			\draw[line width=0.8 pt, color=black!30!white] 
			++(90:0.578) ++(30:0.578)++(90:0.578)
			-- ++(90:0.578)
			-- ++(150:0.578)-- ++(90:0.578) -- ++(150:0.578)-- ++(90:0.578) 
			-- ++(30:0.578)--++(90:0.578) -- ++(30:0.578)--++(90:0.578) -- ++(30:0.578)
			--++(-30:0.578) -- ++(30:0.578)--++(-30:0.578) -- ++(30:0.578)--++(-30:0.578)
			-- ++(-90:0.578)--++(-30:0.578) -- ++(-90:0.578)--++(-30:0.578) -- ++(-90:0.578)
			--++(-150:0.578)--++(-90:0.578)--++(-150:0.578)--++(-90:0.578)--++(-150:0.578)
			--++(150:0.578)--++(210:0.578)--++(150:0.578)--++(210:0.578)--cycle;
			
			% letters
			%\wordbox{black}{4}{4}{\large{$\mathfrak{D}_{\text{int}}$}};
			\node[]  at (2.2, 3.4){\large{$\mathfrak{D}_{\text{int}}$}};
			\node[]  at (5.5, 0.5){\large{$\mathfrak{D}_{\partial}$}};
			\draw[->, line width=0.7 pt, color=black] 
			(4.6, 0.9)
			.. controls +(180:0.3) and +(-80:0.4) ..
			(3.9, 1.9);
			\node[]  at (2, -1.3){{(a)}};
		\end{scope}
		
		\begin{scope}[xshift= 4 cm, scale=0.38]
			% 0th line
			\diskshadded{-2}{0}{6}{0}{gray!30!white}{white,opacity=0.4};
			%1st line
			\diskshadded{-1}{1}{0}{1}{gray!30!white}{white,opacity=0.4};
			\diskshadded{1}{1}{4}{1}{blue!50!cyan!30!white}{blue!50!cyan!10!white,opacity=0.4};
			\diskshadded{5}{1}{6}{1}{gray!30!white}{white,opacity=0.4};
			%2nd line
			\diskshadded{-1}{2}{0}{2}{gray!30!white}{white,opacity=0.4};
			\diskshadded{1}{2}{5}{2}{blue!50!cyan!30!white}{blue!50!cyan!10!white,opacity=0.4};
			\diskshadded{6}{2}{7}{2}{gray!30!white}{white,opacity=0.4};
			%3rd line
			\diskshadded{0}{3}{0}{3}{gray!30!white}{white,opacity=0.4};
			\diskshadded{1}{3}{6}{3}{blue!50!cyan!30!white}{blue!50!cyan!10!white,opacity=0.4};
			\diskshadded{7}{3}{7}{3}{gray!30!white}{white,opacity=0.4};
			%4th line
			\diskshadded{0}{4}{0}{4}{gray!30!white}{white,opacity=0.4};
			\diskshadded{1}{4}{7}{4}{blue!50!cyan!30!white}{blue!50!cyan!10!white,opacity=0.4};
			\diskshadded{8}{4}{8}{4}{gray!30!white}{white,opacity=0.4};
			%5th line
			\diskshadded{1}{5}{1}{5}{gray!30!white}{white,opacity=0.4};
			\diskshadded{2}{5}{7}{5}{blue!50!cyan!30!white}{blue!50!cyan!10!white,opacity=0.4};
			\diskshadded{8}{5}{8}{5}{gray!30!white}{white,opacity=0.4};
			%6th line
			\diskshadded{1}{6}{2}{6}{gray!30!white}{white,opacity=0.4};
			\diskshadded{3}{6}{7}{6}{blue!50!cyan!30!white}{blue!50!cyan!10!white,opacity=0.4};
			\diskshadded{8}{6}{9}{6}{gray!30!white}{white,opacity=0.4};
			%7st line
			\diskshadded{2}{7}{3}{7}{gray!30!white}{white,opacity=0.4};
			\diskshadded{4}{7}{7}{7}{blue!50!cyan!30!white}{blue!50!cyan!10!white,opacity=0.4};
			\diskshadded{8}{7}{9}{7}{gray!30!white}{white,opacity=0.4};
			% 8th line
			\diskshadded{2}{8}{10}{8}{gray!30!white}{white,opacity=0.4};
			
			% draw the line
			\draw[line width=0.8 pt] (30:0.578) -- (90:0.578) 
			-- ++(90:0.578)
			-- ++(150:0.578)-- ++(90:0.578) -- ++(150:0.578)-- ++(90:0.578) -- ++(150:0.578)-- ++(90:0.578) 
			-- ++(30:0.578)--++(90:0.578) -- ++(30:0.578)--++(90:0.578) -- ++(30:0.578)--++(90:0.578) -- ++(30:0.578)
			--++(-30:0.578) -- ++(30:0.578)--++(-30:0.578) -- ++(30:0.578)--++(-30:0.578) -- ++(30:0.578)--++(-30:0.578)
			-- ++(-90:0.578)--++(-30:0.578) -- ++(-90:0.578)--++(-30:0.578) -- ++(-90:0.578)--++(-30:0.578) -- ++(-90:0.578)
			--++(-150:0.578)--++(-90:0.578)--++(-150:0.578)--++(-90:0.578)--++(-150:0.578)--++(-90:0.578)--++(-150:0.578)
			--++(150:0.578)--++(210:0.578)--++(150:0.578)--++(210:0.578)--++(150:0.578)--++(210:0.578)--cycle;

			% add terms
			\coloredkv{yellow!60!white,opacity=0.5}{0}{6}{4};
			% K_f
			\coloredkf{green!60!yellow!50!white,opacity=0.5}{0}{1}{1};
			\coloredkf{green!60!yellow!50!white,opacity=0.5}{3}{2}{4};
			% K_e
			\coloredke{yellow!40!red!50!white,opacity=0.5}{1}{6}{6};
			\coloredke{yellow!40!red!50!white,opacity=0.5}{3}{4}{6};
			\coloredke{yellow!40!red!50!white,opacity=0.5}{2}{4}{3};
			
			\node[]  at (2, -1.3){{(b)}};
		\end{scope}
		
	\end{tikzpicture}
	\caption{(a) A disk $\mathfrak{D}$ and its partition into $\mathfrak{D}_{\text{int}}$ and $\mathfrak{D}_{\partial}$. (b) A disk $\mathfrak{D}$ (blue) and the 1-site terms $K_v$ (yellow), 2-site terms $K_e$ (orange), 3 site terms $K_f$ (green) in the local decomposition (\ref{eq:mod_decomp}) of modular Hamiltonian $K_{\mathfrak{D}}$.}
	\label{fig:local_terms}
\end{figure}

To take advantage of this process, let us coarse-grain the subsystem into a triangular
lattice of supersites; see Fig.~\ref{fig:local_terms}(a).  This lattice has \textit{vertices}
$\tikz[scale=0.2,baseline=0ex,yshift=-0.15cm]{
	\diskshadded{1}{1}{1}{1}{black}{white,opacity=0.4};
}$ (denoted by hexagonal unit cells at each site in the figure), edges 
$\tikz[scale=0.2,baseline=0ex,yshift=-0.15cm]{
	\diskshadded{1}{1}{1}{1}{black}{white,opacity=0.4};
	\diskshadded{2}{1}{2}{1}{black}{white,opacity=0.4};
}$, as well as three-site combinations such as faces 
$\tikz[scale=0.2,baseline=0ex,yshift=-0.15cm]{
	\diskshadded{1}{1}{1}{1}{black}{white,opacity=0.4};
	\diskshadded{2}{1}{2}{1}{black}{white,opacity=0.4};
	\diskshadded{1}{0}{1}{0}{black}{white,opacity=0.4};
}$,
$\tikz[scale=0.2,baseline=0ex,yshift=-0.15cm]{
	\diskshadded{2}{1}{2}{1}{black}{white,opacity=0.4};
	\diskshadded{1}{0}{1}{0}{black}{white,opacity=0.4}; 
	\diskshadded{2}{0}{2}{0}{black}{white,opacity=0.4};
}$
and chain-like regions
$\tikz[scale=0.2,baseline=-0.5ex,yshift=0.32cm]{
	\diskshadded{0}{0}{2}{0}{black}{white,opacity=0.4};
}$,
$\tikz[scale=0.2,baseline=0ex,yshift=-0.15cm]{
	\diskshadded{1}{1}{2}{1}{black}{white,opacity=0.4};
	\diskshadded{0}{0}{0}{0}{black}{white,opacity=0.4};
}$, $\dots$
The above procedure eliminates all terms supported on chain-like regions, leaving
the faces $f\in F(\mathfrak{D})$ as the only source of three-site supports for the decomposition. 
The decomposition also admits single-site terms, which,
due to the nature of Eq.~\eqref{eq:markov_decomposition}, only
occupy vertices $v$ on the interior $\mathfrak{D}_{\text{int}}$, \emph{i.e.,} the set of all vertices that have no neighbors
outside of $\mathfrak{D}$. Similarly, the remaining two-site terms
are supported on any edges $e\in E(\mathfrak{D})$ that do not lie
exclusively on the boundary, \emph{i.e.,} the set $E(\mathfrak{D})/E(\mathfrak{D}_{\partial})$.
These terms make up the decomposition
\begin{equation}
K_{\mathfrak{D}}=\sum_{\substack{f\in F(\mathfrak{D})}
}K_{f}-\sum_{e\in E(\mathfrak{D})/E(\mathfrak{D}_{\partial})}K_{e}+\sum_{v\in\mathfrak{D}_{\text{int}}}K_{v}~,\label{eq:mod_decomp}
\end{equation}
which is local with respect to our triangulation; examples of each
term are depicted in Fig.~\ref{fig:local_terms}(b).

Invariance of $J(A,B,C)$ under smooth deformations implies that commutators of the above terms are $0,\pm J$, where we have set 
\begin{equation}\label{eq:alpha_def}
    J:=J(u,v,w)\quad\text{on a face of the form}\quad    \tikz[scale=0.4,baseline=5ex]{	
		\coloredboxhx{white,opacity=0.5}{6}{2}{7}{2};
			\coloredboxhx{white,opacity=0.5}{7}{3}{7}{3};
			\node at (5,1.73) {$w$};
			\node at (5.5,2.55) {$v$};
			\node at (6,1.73) {$u$};
	}~. 
\end{equation}
This reference value of the modular commutator either stays the same or changes sign if we pick another face or rearrange the order of the sites.

\emph{Modular current---}Because $K_\mathfrak{D}$ is local, we can define its ``energy current,'' which we refer to as the modular current. 
We rewrite the modular Hamiltonian~\eqref{eq:mod_decomp} in terms of $\widetilde{K}^{\mathfrak{D}}_v$, which collects all terms whose support contains site $v$, multiplied by the appropriate fraction so as to satisfy $K_{\mathfrak{D}} = \sum_{v\in \mathfrak{D}} \widetilde{K}^{\mathfrak{D}}_v$:
\begin{equation}\label{eq:mod_decomp2_addendum}
\widetilde{K}_{v}^{\mathfrak{D}}=
\begin{cases}
{\displaystyle {\textstyle\frac{1}{3}}\sum_{f:v\in f}K_{f}-{\textstyle\frac{1}{2}}\sum_{e:v\in e}K_{e}+K_{v}}, & v\in\mathfrak{D}_{\text{int}}\\
{\displaystyle {\textstyle\frac{1}{3}}\sum_{\substack{f:v\in f,\\
f\in F(\mathfrak{D})
}
}K_{f}-{\textstyle\frac{1}{2}}\sum_{\substack{e:v\in e,\\
e\in E(\mathfrak{D})\setminus E(\mathfrak{D}_{\partial})
}
}K_{e}}, & v\in\mathfrak{D}_{\partial}
\end{cases}
~.
\end{equation}
The modular current from site $u$ to $v$ is then simply
\begin{equation}
	f^{\mathfrak{D}}_{uv} := i\langle [\widetilde{K}^{\mathfrak{D}}_u, \widetilde{K}^{\mathfrak{D}}_v]\rangle~,
\end{equation}
quantifying the non-commutativity of the modular Hamiltonian terms.
 
The modular current has properties analogous to that of the energy current of some local Hamiltonian at finite temperature~\cite{Kitaev2006a}. For instance, $f^{\mathfrak{D}}_{uv}$ vanishes for any pair of points $u, v\in {\mathfrak{D}}$ which are sufficiently far apart. Moreover, the current is conserved:
\begin{equation}
    	\sum_{v\in \mathfrak{D}} f^{\mathfrak{D}}_{uv}=0. \label{eq:conservation_law}
\end{equation}
Lastly, one can explicitly show that the bulk modular current vanishes: $f^{\mathfrak{D}}_{uv} =0$ for  $u,v\in \mathfrak{D}_{\text{int}}$; see the Supplemental Material for details.

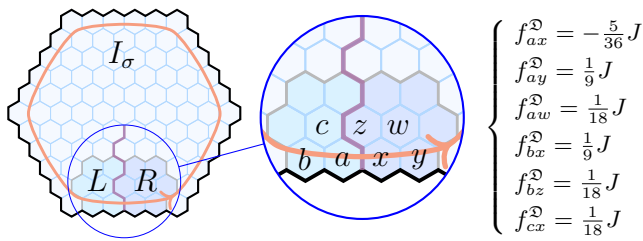
\begin{figure}[h]
	\centering
	\begin{tikzpicture}
		\begin{scope}[scale=0.8]
		\begin{scope}[scale=0.35,spy using outlines={circle, magnification=1.8, size=1 cm, connect spies}]
			
			% recolor for L and R (newly added)
			\diskshadded{1}{1}{6}{1}{blue!50!cyan!30!white}{blue!50!cyan!10!white,opacity=0.4};
			    \diskshadded{2}{1}{3}{1}{blue!50!cyan!30!white}{blue!30!cyan!30!white,opacity=0.4};
			    \diskshadded{4}{1}{5}{1}{blue!50!cyan!30!white}{blue!70!cyan!30!white,opacity=0.4};
			\diskshadded{1}{2}{7}{2}{blue!50!cyan!30!white}{blue!50!cyan!10!white,opacity=0.4};
			    \diskshadded{2}{2}{3}{2}{blue!50!cyan!30!white}{blue!30!cyan!30!white,opacity=0.4};
			    \diskshadded{4}{2}{6}{2}{blue!50!cyan!30!white}{blue!70!cyan!30!white,opacity=0.4};
			\diskshadded{1}{3}{8}{3}{blue!50!cyan!30!white}{blue!50!cyan!10!white,opacity=0.4};
			    \diskshadded{3}{3}{4}{3}{blue!50!cyan!30!white}{blue!30!cyan!30!white,opacity=0.4};
			    \diskshadded{5}{3}{6}{3}{blue!50!cyan!30!white}{blue!70!cyan!30!white,opacity=0.4};
			\diskshadded{1}{4}{9}{4}{blue!50!cyan!30!white}{blue!50!cyan!10!white,opacity=0.4};
			  %  \diskshadded{4}{4}{4}{4}{blue!50!cyan!30!white}{blue!30!cyan!30!white,opacity=0.4};
			   % \diskshadded{5}{4}{6}{4}{blue!50!cyan!30!white}{blue!70!cyan!30!white,opacity=0.4};
			\diskshadded{1}{5}{10}{5}{blue!50!cyan!30!white}{blue!50!cyan!10!white,opacity=0.4};
			\diskshadded{1}{6}{11}{6}{blue!50!cyan!30!white}{blue!50!cyan!10!white,opacity=0.4};
			
			\diskshadded{2}{7}{11}{7}{blue!50!cyan!30!white}{blue!50!cyan!10!white,opacity=0.4};
			\diskshadded{3}{8}{11}{8}{blue!50!cyan!30!white}{blue!50!cyan!10!white,opacity=0.4};
			\diskshadded{4}{9}{11}{9}{blue!50!cyan!30!white}{blue!50!cyan!10!white,opacity=0.4};
			\diskshadded{5}{10}{11}{10}{blue!50!cyan!30!white}{blue!50!cyan!10!white,opacity=0.4};
			\diskshadded{6}{11}{11}{11}{blue!50!cyan!30!white}{blue!50!cyan!10!white,opacity=0.4};

			% added boundry of L and R
		\draw[line width=0.6 pt, color=gray!55!white] (30:0.578) -- ++(30:0.578) 
		-- ++(90:0.578) -- ++(150:0.578)-- ++(90:0.578) -- ++(30:0.578)-- ++(90:0.578) -- ++(30:0.578) -- ++(-30:0.578)-- ++(30:0.578)
		-- ++(-30:0.578)-- ++(30:0.578)-- ++(-30:0.578)-- ++(30:0.578)
		--++(-30:0.578)--++(-90:0.578)--++(-30:0.578)--++(-90:0.578)--++(-150:0.578)--++(-90:0.578);
			
			\draw[line width=0.8 pt] (30:0.578) -- (90:0.578) 
			-- ++(90:0.578) -- ++(150:0.578)-- ++(90:0.578) -- ++(150:0.578)-- ++(90:0.578) -- ++(150:0.578)-- ++(90:0.578) -- ++(150:0.578)-- ++(90:0.578) -- ++(150:0.578)
			--++(90:0.578) -- ++(30:0.578)--++(90:0.578) -- ++(30:0.578)--++(90:0.578) -- ++(30:0.578)--++(90:0.578) -- ++(30:0.578)--++(90:0.578) -- ++(30:0.578)--++(90:0.578) -- ++(30:0.578)
			--++(-30:0.578) -- ++(30:0.578)--++(-30:0.578) -- ++(30:0.578)--++(-30:0.578) -- ++(30:0.578)--++(-30:0.578) -- ++(30:0.578)--++(-30:0.578) -- ++(30:0.578)--++(-30:0.578) 
			-- ++(-90:0.578)--++(-30:0.578) -- ++(-90:0.578)--++(-30:0.578) -- ++(-90:0.578)--++(-30:0.578) -- ++(-90:0.578)--++(-30:0.578) -- ++(-90:0.578)--++(-30:0.578) 
			--++(-90:0.578)--++(-150:0.578)--++(-90:0.578)--++(-150:0.578)--++(-90:0.578)--++(-150:0.578)--++(-90:0.578)--++(-150:0.578)--++(-90:0.578)--++(-150:0.578)--++(-90:0.578)--++(-150:0.578)
			--++(150:0.578)--++(210:0.578)--++(150:0.578)--++(210:0.578)--++(150:0.578)--++(210:0.578)--++(150:0.578)--++(210:0.578)--++(150:0.578)--++(210:0.578)--cycle;

		\draw[line width=1 pt,color=purple!70!blue!75!black!55!white, opacity=0.8] (30:0.578) ++ (0:2)++(30:0.578)--++ (90:0.578)--++(150:0.578)--++(90:0.578)--++(30:0.578)--++(90:0.578)--++(150:0.578)--++(90:0.578)--++(30:0.578)--++(90:0.578) ;
		\begin{scope}[xshift=3 cm, yshift=5.18 cm]
		%\draw[line width=0.8 pt] (0:0)--(4.5,0);
		\draw[->, line width=1.1 pt, color=red!50!orange!93!black!50!white] 
		(-60:4.55) .. controls +(30:0.5) and +(-120:0.5) ..
        (-30:4.25) .. controls +(60:0.5) and +(-90:0.5) ..       
		(0:4.55) .. controls +(90:0.5) and +(-60:0.5) ..
		(30:4.25) .. controls +(120:0.5) and +(-30:0.5) ..		
		(60:4.55) .. controls +(150:0.5) and +(0:0.5) ..
		(90:4.25) .. controls +(180:0.5) and +(30:0.5) ..		
		(120:4.55) .. controls +(210:0.5) and +(60:0.5) ..
		(150:4.25) .. controls +(240:0.5) and +(90:0.5) ..		
		(180:4.55) .. controls +(270:0.5) and +(120:0.5) ..
		(210:4.25) .. controls +(300:0.5) and +(150:0.5) ..		
		(240:4.55) .. controls +(330:0.5) and +(180:0.5) ..
		(270:4.25) .. controls +(0:0.5) and +(210:0.5) ..
		(301:4.55);
		\end{scope}
		\node[]  at (2.9, 8){\large{${I_{\sigma}}$}};

	\spy [blue, size=2.7 cm] on (0.84, 0.55) in node [] at (14.3,5.0);

\end{scope}	
\begin{scope}
		% add labels L and R
	\node[]  at (1.4,0.7){\large{${R}$}};
	\node[]  at (0.6, 0.7){\large{${L}$}};
\end{scope}

\begin{scope}[scale=0.35, xshift=18.2 cm, yshift=8.4 cm, scale=0.91]	
		\node[]  at (-7.5+0.2,-5.9) {\large{$b$}};
	\node[]  at (-5.5+0.2,-5.9) {\large{$a$}};
	\node[]  at (-6.6+0.2,-5.9+0.578+1.1) {\large{$c$}};
	
	\node[]  at (-3.5+0.2,-5.9) {\large{$x$}};
	\node[]  at (-1.5+0.2,-5.9) {\large{$y$}};
	\node[]  at (-4.6+0.2,-5.9+0.578+1.1) {\large{$z$}};
	\node[]  at (-2.6+0.2,-5.9+0.578+1.1) {\large{$w$}};

	% numbers
	\begin{scope}[xshift=-4cm,yshift=0.25 cm]
	\node[]  at (10.5,-4.5) {$\left\{\begin{array}{l}
			f^{\mathfrak{D}}_{ax}=-\frac{5}{36}J\\[3pt]
			f^{\mathfrak{D}}_{ay}=\frac{1}{9}J\\[3pt]
			f^{\mathfrak{D}}_{aw}=\frac{1}{18}J\\[3pt]
			f^{\mathfrak{D}}_{bx}=\frac{1}{9}J\\[3pt]
			f^{\mathfrak{D}}_{bz}=\frac{1}{18}J\\[3pt]
			f^{\mathfrak{D}}_{cx}=\frac{1}{18}J		
		\end{array}\right.$};
\end{scope}
\end{scope}	
\end{scope}
	\end{tikzpicture}
	\caption{The calculation of the edge modular current $I_{\sigma}$~\eqref{eq:edge_current_value}. The computation result of all the non-vanishing contribution from modular currents passing through a specific cut is summarized (see the Supplemental Material for details).} 
	\label{fig:edge_modular_current}
\end{figure}

Since the modular current vanishes in the bulk of the disk, nontrivial current flows only along the edge. We define the \emph{edge modular current} $I_\sigma$ to be a sum of $f_{uv}^{\mathfrak{D}}$ over $u\in{L}$ and $v\in{R}$, where ${L},{R}\subset\mathfrak{D}$ lie on opposite sides of a cut perpendicular to the boundary. This current is insensitive to the choice of regions ${L},{R}$, as long as they are sufficiently large. Since there is no current deep enough in the bulk, nonzero contributions only come from sites that are at most two sites away from the edge. Moreover, because the current flows between nearby sites and is conserved, it suffices to consider only a few sites along the edge. For the cut depicted in Fig.~\ref{fig:edge_modular_current}, nonzero contributions only come from $\{a,b,c\} \subset {L}$ and $\{x,y,z,w\} \subset {R}$, yielding:
\begin{equation}
	I_{\sigma}=\sum_{u\in{L}}\sum_{v\in{R}}f_{uv}^{\mathfrak{D}}=\frac{1}{4}J~.\label{eq:edge_current_value}
\end{equation}
The choice of regions depends on the decomposition of $K_\mathfrak{D}$, and a coarser decomposition requires even smaller regions.

\emph{Chiral central charge---}We now relate $I_{\sigma}$ to $c_-$ by viewing $K_{\mathfrak{D}}$ as a physical Hamiltonian describing the same phase as that of the state $\sigma$.
First, $\sigma_{\mathfrak{D}}=e^{-K_{\mathfrak{D}}}$ can be viewed as a thermal state of the local Hamiltonian $K_{\mathfrak{D}}$ with the temperature $T=1$.  Second,  $\sigma_{\mathfrak{D}}$ obeys an entanglement area law in $\mathfrak{D}$, and is indistinguishable from the ground state $\sigma$ over the same region, an indication of \emph{low} temperature for a system with a bulk energy gap. Third, the temperature is \emph{high enough} so that the edge correlation length is small compared to the length of the disk. This is the right temperature range for which we can apply Eq.~\eqref{eq:edge_current}~\cite{Kitaev2006a}. 

Now, we can make a nontrivial but a reasonable physical assumption; if two local Hamiltonians with a bulk energy gap have the same bulk reduced density matrix, then their energy currents at the boundary at a temperature low compared to the bulk gap both obey Eq.~\eqref{eq:edge_current}, independent of other microscopic details. This expression is determined completely by the temperature and the CCC of the edge theory~\cite{Kitaev2006a} and cannot change unless the bulk undergoes a quantum phase transition. As is observed in the previous paragraph, the ``temperature'' $T=1$  lies in the right temperature range. Using Eq.~\eqref{eq:edge_current} and identifying the energy current to the edge modular current~\eqref{eq:edge_current_value}, we arrive at our main result:
\begin{equation}
    J=\frac{\pi}{3} c_-. \label{eq:main2}
\end{equation}

\emph{Discussion---}In this Letter, we propose a new formula for the chiral central charge (CCC). The formula is based on the modular commutator $J$, a new quantity that is expressed in terms of reduced density matrices and that is odd under time reversal (i.e., complex conjugation with respect to a local basis). Compared to other similar works in this direction~\cite{Zhang2012,Tu2013,Bradlyn2015a}, the main advantage of our formula is that it can be directly computed from a single ground state wave function. Together with the prior work that established a connection between ground state entanglement and the anyonic data~\cite{Kitaev2006,Levin2006,Li2008,Zhang2012,SKK2019,Shi2020b}, our result strengthens the idea that \emph{all} the universal properties of $(2+1)$-dimensional gapped quantum phases may be encoded in a single ground state. Our formula is thus a useful addition to the existing toolkit for diagnosing topological properties of quantum many-body systems. 

Aside from drawing a deep connection between ground state entanglement and the CCC, our result~\eqref{eq:main2} also has an intriguing implication for the \emph{intrinsic sign problem}~\cite{Hastings2016}. Specifically, if our formula for the CCC is correct for all gapped systems in (2+1) dimensions, then a system with nonzero CCC cannot have a parent Hamiltonian which is sign-problem free. 
This is because a sign-problem free Hamiltonian necessarily admits a ground state with nonnegative coefficients in a local product basis, for which the CCC would be zero because $J$ is odd under complex conjugation. Thus, a rigorous proof of our formula will lead to a microscopic proof that such systems have an intrinsic sign problem; no local Hamiltonian free of the sign problem can have ground states with nonzero $c_-$. This corroborates recent arguments pointing to a similar conclusion~\cite{Ringele2017,Golan2020,Golan2021}. 

A curious fact is that the modular commutator did not possess any ultraviolet-divergent contributions in our calculation. Whether this is a general property of modular commutator or something specific to gapped phases in $(2+1)$ dimensions is unclear. Perhaps field-theoretic calculations of our result can shed light on this problem.

For future work, it will be interesting to find analogues of $J$ that are suitable for resolving symmetry-protected topological invariants that can be obtained from the ground state wave function, \emph{e.g.,}
Refs.~\cite{Fidkowski2011,Turner2011,Haegeman2012,Marvian2013,Shiozaki2017,Shapourian2017,Tasaki2018,Shiozaki2018,Bachmann2020,Cian2021,Dehghani2021}.

\subsection*{Acknowledgments:}
The authors thank Nikita Sopenko, Michael Gullans, Tarun Grover and 
Maissam Barkeshli for valuable input and inspiring discussions. I.K. was supported by the Australian Research Council via the Centre of Excellence in Engineered Quantum Systems (EQUS) project number CE170100009. B.S. would like to thank John McGreevy and Jin-Long Huang for being extremely supportive when B.S. periodically takes time off from another project due to the fascination of Isaac's new idea. B.S. is supported by University of California Laboratory Fees Research Program, grant LFR-20-653926, and the Simons Collaboration on Ultra-Quantum Matter, grant 651440 from the Simons Foundation. K.K. is supported by MEXT Quantum LeapFlagship Program (MEXT Q-LEAP) Grant Number JP-MXS0120319794. Contributions to this work by NIST, an agency of the US government, are not subject to US copyright. Any mention of commercial products does not indicate endorsement by NIST. V.V.A. thanks Olga Albert, Halina and Ryhor Kandratsenia, as well as Tatyana and Thomas Albert for providing daycare support throughout this work.

\bibliography{bib,vva}

\appendix

\section{Fermionic Markov chain}
\label{appendix:fermion}
For fermions, the global Hilbert space does not have a tensor product structure. Thus, one must define a proper notion of partial trace more precisely. We provide a self-contained overview of this subject, following the treatment of Ref.~\cite{Araki2003,Nacthergaele2017a}. We focus on fermionic systems on a finite lattice, which removes the subtleties that may appear in infinite systems. 

The fermionic algebra is defined in terms of the canonical anticommutation relation:
\begin{equation}
    \{a_j^{\dagger}, a_k \} = \delta_{j,k},\quad
    \{a_j, a_k \}=\{a_j^{\dagger}, a^{\dagger}_k \} = 0, \label{eq:anticomm}
\end{equation}
where $a_j^{\dagger}$ and $a_j$ are the fermionic creation and annihilation operators, respectively, and $\Lambda \ni j$ is the set of fermion modes. The algebra generated by these operators is denoted as $\mathcal{A}_\Lambda$. More generally, we denote a subalgebra generated by $\{a_j^{\dagger}, a_j:  j \in A\subset \Lambda \}$ as $\mathcal{A}_A$.

To define a proper notion of partial trace, it will be helpful to consider a slight modification of the partial trace for qubits that admits a generalization in a straightforward way. Let $\Lambda'$ be a set of qubits. A partial trace over a subset $A' \subset \Lambda'$ can be formally defined as follows:
\begin{equation}
\begin{aligned}
\rho_{\Lambda'} &\to    \frac{1}{4^{|A'|}} \sum_{P_{A'}} P_{A'} \rho P_{A'}^{\dagger} \\
&= \frac{I_{A'}}{2^{|A'|}} \otimes \rho_{\bar{A'}} 
\end{aligned}
\label{eq:modified_partial_trace}
\end{equation}
where the summation is taken over the $4^{|A'|}$ Pauli strings acting on $A'$ (i.e., a product of single site Pauli operators and the identity operator $\{I_j, X_j,Y_j,Z_j\}$ with $j \in A'$)  and $I_{A'}$ is the identity matrix acting on $A'$. $\bar{A'}= \Lambda'\setminus A'$ is the complement of $A'$.

There is a simple analog of Eq.~\eqref{eq:modified_partial_trace} for fermions, as explained in Ref.~\cite{Nacthergaele2017a}.  Density matrices of fermionic systems are operators $\rho_{\Lambda} \in \mathcal{A}_{\Lambda}^{+}$, such that $\rho_{\Lambda}$ is positive. Here $\mathcal{A}_{\Lambda}^{+} \subset \mathcal{A}_{\Lambda}$ is the subalgebra formed by terms written as linear combination of even powers of the creation and annihilation operators.

Consider the following unitary operators:
\begin{equation}
\begin{aligned}
    u_j^{(0)} &= I \\
    u_j^{(1)} &= a_j^{\dagger} + a_j \\
    u_j^{(2)} &= a_j^{\dagger} - a_j \\
    u_j^{(3)} &= I - 2a_j^{\dagger}a_j.
\end{aligned}
\end{equation}
The fermionic analog of Eq.~\eqref{eq:modified_partial_trace} is
	\begin{equation}
    \fTr_A (\rho_{\Lambda}) \equiv \frac{1}{4^{|A|}} \sum_{\alpha \in \text{Ind}_{A}} u(\alpha)^{\dagger} \rho_{\Lambda} u(\alpha),
\end{equation}
where $\text{Ind}_{A} = \{0,1,2,3 \}^{|A|}$ and $u(\alpha) = u_{j_1}^{(\alpha_1)}\ldots u_{j_n}^{(\alpha_n)}$ for $A = \{j_1,\ldots, j_n \}$. Let us note that the ordering of $u_{j_n}^{(\alpha_n)}$ is immaterial because they either commute or anticommute with each other.

It is straightforward to verify that $\fTr_A$ satisfies the following intuitive properties of partial trace:
\begin{enumerate}
    \item $\fTr_A(O) = O$ for $O\in \mathcal{A}_{\Lambda \setminus A}^{+}$.
    \item $\fTr_A \circ \fTr_B = \fTr_B \circ \fTr_A= \fTr_{AB}$, for $A\cap B= \emptyset$.
    \item $\fTr_A(O) \in \mathcal{A}_{\Lambda \setminus A}^{+}$ for for $O\in \mathcal{A}_{\Lambda}^{+}$.
\end{enumerate}
Thus, it is sensible to define the following notion of reduced density matrix for fermions:
\begin{equation}\label{eq:f_rho_A}
    \widetilde{\rho}_A = \fTr_{\bar{A}}(\rho).
\end{equation}
Armed with this knowledge, one can prove a necessary and sufficient condition for saturation of strong subadditivity for fermions.

\subsubsection{Relating different reduced density matrices}
Now we relate $\widetilde{\rho}_A$ defined in Eq.~(\ref{eq:f_rho_A}) to a more familiar notion of reduced density matrix. This is done by associating a linear functional (trace) which maps operators in the algebra $\mathcal{A}_{\Lambda}$ (or $\mathcal{A}_{A}$) to complex numbers. 

This is most convenient if we introduce the Majorana operators $\{ c_j, d_j \}_{j\in \Lambda}$ such that $c_j=u_j^{(1)}$ and $d_j=i u_j^{(2)}$, and notice that the algebra $\mathcal{A}_{\Lambda}$ is generated $\{ c_j, d_j \}_{j\in \Lambda}$. In other words, every element in $\mathcal{A}_{\Lambda}$ is a linear combination of monomials of the form
$	\prod_{j \in \Lambda} (c_j^{s_j} d_j^{t_j}) $,
where $s_j, t_j \in \{ 0,1\}$.

We formally define the trace operation (for operators in  $\mathcal{A}_{\Lambda}$)  according to
\begin{equation}\label{eq:trace_lambda}
\Tr \,	\prod_{j \in \Lambda} (c_j^{s_j} d_j^{t_j})\equiv 2^{|\Lambda |}  \prod_{j\in \Lambda} (\delta_{s_j,0} \cdot \delta_{t_j,0}).
\end{equation}
(Because the trace is a linear operation, the properties on the monomials above completely determine the trace operation.)

Now we look back at the operator $\widetilde{\rho}_A$ defined in Eq.~(\ref{eq:f_rho_A}). Because $\widetilde{\rho}_A \in \mathcal{A}_{A}^{+}\subset \mathcal{A}_{\Lambda}$, we have implicitly chosen the trace defined on $\mathcal{A}_{\Lambda}$, according to  Eq.~(\ref{eq:trace_lambda}). However, since $\mathcal{A}_{A}^{+}\subset \mathcal{A}_{A}$, we can choose a different normalization of the trace, defined according to:
\begin{equation}\label{eq:trace_A}
	\Tr^{(A)} \,	\prod_{j \in A} (c_j^{s_j} d_j^{t_j})\equiv 2^{|A|}  \prod_{j\in A} (\delta_{s_j,0} \cdot \delta_{t_j,0}).
\end{equation}
To distinguish the two traces, we introduced an upper index $(A)$.
Intuitively, this different normalization is possible because the Hilbert space associated with ${\Lambda}$ decomposes into $2^{|\Lambda|- |A|}$ identical copies of irreducible representation of the algebra $\mathcal{A}_{A}$. 

Let $\rho_A \in \mathcal{A}^{+}_{A}$ be a density matrix in terms of the trace defined in Eq.~(\ref{eq:trace_A}), obtained by rescaling $\widetilde{\rho}_A$, then the normalization $\Tr^{(A)} \rho_A =1$ (instead of $\Tr \rho_A =1$) determines that
\begin{equation}
	\widetilde{\rho}_A =\rho_A / 2^{|\Lambda\setminus A|}.
\end{equation}
In particular, this implies that 
\begin{equation}
    \ln \widetilde{\rho}_A = \ln \rho_A - I(|\Lambda| - |A|) \ln 2.
\end{equation}
Thus,
\begin{equation}
\begin{aligned}
    \ln \widetilde{\rho}_{AB} +  \ln \widetilde{\rho}_{BC} -  \ln \widetilde{\rho}_{B} -  \ln \widetilde{\rho}_{ABC} 
    \\
    =
    \ln\rho_{AB} +  \ln \rho_{BC} - \ln\rho_{B} -  \ln \rho_{ABC},
\end{aligned}
\label{eq:operator_cmi_identity}
\end{equation}
and
\begin{equation}
	\Tr(\widetilde{\rho}_{A} \ln \widetilde{\rho}_{A}) =	\Tr^{(A)} ({\rho}_{A} \ln {\rho}_{A})- (|\Lambda| -|A|) \ln 2. 
\end{equation}
The latter condition implies that $I(A:C|B)_{\rho}=I(A:C|B)_{\widetilde{\rho}}$.

\subsubsection{Explicit basis}
The discussion above took an algebraic approach, and the analysis did not need to specify any basis of the Hilbert space. For practical application it is often useful to think of   the standard fermionic occupation number basis, with basis vectors $\{ \vert n_1, \cdots, n_{\Lambda} \rangle \}$, defined according to
\begin{equation}
	a_1^{\dagger n_1}\cdots a_{|\Lambda|}^{\dagger n_{|\Lambda|}} |0,\cdots,0\rangle =|n_1,\cdots, n_{|\Lambda|}\rangle,
\end{equation}
where $n_j \in \{0,1\}$. Note that the ordering in the product of creation operators matters, and here we have chosen an ordering that increases the site labels.

Now the trace operation looks more standard, in the sense that  the previous defining property of ``$\Tr$" in Eq.~(\ref{eq:trace_lambda}) now becomes the consequence of the formula
\begin{equation}
	\Tr(O)= \sum_I \langle I | O | I\rangle ,\quad \forall O \in \mathcal{A}_{\Lambda}.
\end{equation}
Here $\{ |I \rangle\}$ is a basis set written in terms of the fermionic occupation numbers, \emph{i.e.,} $|I\rangle = |n_1^{(I)}, \ldots, n_{|\Lambda|}^{(I)}\rangle$ where $n_k^{(I)} \in \{ 0, 1\}$. (A similar analysis applies to $\Tr^{(A)}$.)

As an explicit example, let $\Lambda=AB$, where $A$ and $B$ each contains one site. For this system, the total Hilbert space is $4$ dimensional. The algebra $\mathcal{A}_{\Lambda}$ is a 16 dimensional vector space and $\mathcal{A}_{A}$ is 4 dimensional. Consider density matrix $\rho_{AB}= \vert 0,0\rangle \langle 0,0 \vert $ in the fermionic occupation number basis. Then one can easily verify:
\begin{equation}
	\begin{aligned}
		\rho_{AB} &= \frac{1 -i c_1d_1 -i c_2d_2 - c_1d_1c_2d_2}{4}\\
		\widetilde{\rho}_{A}&= \frac{1- i c_1 d_1}{4}\\
		&= \frac{\vert 0,0\rangle \langle 0,0 \vert +\vert 0,1\rangle \langle 0,1 \vert}{2}\\
	\end{aligned}
\end{equation}
and
\begin{equation}
			\rho_A =\frac{1- i c_1 d_1}{2}
	= \vert 0 \rangle \langle 0 \vert.
\end{equation}

\subsubsection{Saturation of strong subadditivity}

Lieb and Ruskai proved the strong subadditivity of entropy~\cite{Lieb1973}:
\begin{equation}
    I(A:C|B)_{\rho} \geq 0. \label{eq:ssa_temp}
\end{equation}
Moreover, Petz showed that, provided that $\rho_{ABC}$ is positive definite, the following equation holds~\cite{Petz2003}:
\begin{equation}\label{eq:Petz_appendix}
 I(A:C \vert B)_{\rho}=0 \quad \Rightarrow \quad   K_{ABC} = K_{AB} + K_{BC} - K_B.
\end{equation}
The main purpose of this section is to prove the statements analogous to Eq.~\eqref{eq:ssa_temp} and~\eqref{eq:Petz_appendix} for fermions. 

The results in this section is well-known but the following proof is new to the best of our knowledge. The key ingredient is Lieb's three-operator generalization of the Golden-Thompson inequality~\cite{Lieb1973a}:
\begin{equation}
    \text{Tr}(e^{\ln X - \ln Y + \ln Z}) \leq \int_{0}^{\infty}\text{Tr}(X (Y+tI)^{-1} Z(Y+tI)^{-1}) dt,\label{eq:Lieb_triple}
\end{equation}
where $I$ here is the identity and $X,Y,Z$ are positive definite operators. 

To complete the proof, we first note a useful condition written in terms of
relative entropy between a pair of density matrices $D(\rho\| \lambda) = \text{Tr}(\rho(\ln \rho - \ln \lambda))$:
\begin{equation}
	I(A:C|B)_{\rho} = D(\widetilde{\rho}_{ABC} \| \widetilde{\sigma}_{ABC}) - \ln(\text{Tr}(\eta_{ABC})), \label{eq:temp1}
\end{equation}
where $\widetilde{\sigma}_{ABC} = \eta_{ABC} / \text{Tr}(\eta_{ABC})$, is the normalized version of $\eta_{ABC} \equiv \exp(\ln \widetilde{\rho}_{AB} + \ln \widetilde{\rho}_{BC} - \ln \widetilde{\rho}_B)$. By Eq.~\eqref{eq:Lieb_triple}, 
\begin{widetext}
\begin{subequations}
\begin{align}
    \text{Tr}\left(\exp(\ln \widetilde{\rho}_{AB} - \ln \widetilde{\rho}_B + \ln \widetilde{\rho}_{BC}  )\right) &\leq \int_{0}^{\infty} \text{Tr}\left( \widetilde{\rho}_{AB} \frac{1}{\widetilde{\rho}_B + tI} \widetilde{\rho}_{BC} \frac{1}{\widetilde{\rho}_B+tI} \right) dt  \nonumber\\
    &= \int_{0}^{\infty} \text{Tr}\left(\fTr_A \circ \fTr_C\left(\widetilde{\rho}_{AB} \frac{1}{\widetilde{\rho}_B + tI} \widetilde{\rho}_{BC} \frac{1}{\widetilde{\rho}_B+tI} \right)\right) dt \nonumber\\
    &= \int_{0}^{\infty} \text{Tr}\left(\fTr_A \left(\widetilde{\rho}_{AB} \frac{1}{\widetilde{\rho}_B + tI}   \fTr_C(\widetilde{\rho}_{BC}) \frac{1}{\widetilde{\rho}_B+tI} \right) \right) dt \nonumber\\
    &= \int_{0}^{\infty} \text{Tr}\left(\fTr_A \left(\widetilde{\rho}_{AB} \frac{\widetilde{\rho}_B}{(\widetilde{\rho}_B + tI)^2} \right) \right) dt \nonumber\\
    &= \int_{0}^{\infty} \text{Tr}\left(\fTr_A \left(\widetilde{\rho}_{AB}\right) \frac{\widetilde{\rho}_B}{(\widetilde{\rho}_B + tI)^2}  \right) dt \nonumber\\
    &= \int_{0}^{\infty} \text{Tr} \frac{\widetilde{\rho}_{B}^2}{(\widetilde{\rho}_B+tI)^2}  dt \nonumber\\
     &=1. \nonumber
\end{align}
\end{subequations}
(In this calculation we have used the fact that $u(\alpha)$ commutes with elements in $\mathcal{A}_{\Lambda}^{+}$.)
\end{widetext}

Therefore, the second term on the right hand side of Eq.~\eqref{eq:temp1} is nonnegative. The first term is nonnegative by the property of the relative entropy. Therefore, if the left hand side of Eq.~\eqref{eq:temp1} is zero, both terms on the right hand side must be zero. The first term is zero if and only if the two arguments are equal. Therefore, we conclude that
\begin{equation}
\ln \widetilde{\rho}_{ABC} = \ln \widetilde{\rho}_{AB} + \ln \widetilde{\rho}_{BC} - \ln \widetilde{\rho}_B. \label{eq:decomp_fermions}
\end{equation} 
Using Eq.~\eqref{eq:operator_cmi_identity}, we conclude that
\begin{equation}
\ln\rho_{ABC} = \ln \rho_{AB} + \ln \rho_{BC} - \ln \rho_B. 
\end{equation} 
In summary, the fermionic version of Eq.~(\ref{eq:Petz_appendix}) and the strong subadditivity hold.

\section{Calculation details: modular current}

Consider the decomposition of the modular Hamiltonian of a disk-shaped region $\mathfrak{D}$, $K_{\mathfrak{D}} = \sum_{v\in \mathfrak{D}} \widetilde{K}^{\mathfrak{D}}_v$, where
\begin{equation}\label{eq:mod_decomp2_appendix}
	\begin{aligned}
		\widetilde{K}^{\mathfrak{D}}_v &= \frac{1}{3}\sum_{f: v\in f} K_f - \frac{1}{2}\sum_{e:v\in e} K_e + K_v,\quad \forall \,v\in \mathfrak{D}_{\text{int}}\\
		\widetilde{K}^{\mathfrak{D}}_v &= \frac{1}{3}\sum_{\substack{f: v\in f,\\f\in F(\mathfrak{D})}} K_f - \frac{1}{2}\sum_{\substack{e:v\in e,\\ e\in E(\mathfrak{D}) \setminus E(\mathfrak{D}_{\partial})}} K_e,\quad 	\forall \, v\in \mathfrak{D}_{\partial}.
	\end{aligned}%\nonumber
\end{equation}
We calculate the modular current explicitly in this decomposition below.\footnote{Note that the modular current depends on the decomposition of the modular Hamiltonian. However, we can expect the edge current to be insensitive to such choices as long as the bulk current vanishes.} Recall the definition of the modular current from $v$ to $u$:
\begin{equation}
	f^{\mathfrak{D}}_{vu} := i\langle [\widetilde{K}^{\mathfrak{D}}_v, \widetilde{K}^{\mathfrak{D}}_u]\rangle.
\end{equation}

\subsubsection{Vanishing of modular current on $\mathfrak{D}_{\text{int}}$}
\begin{figure}[h]
	\centering	
	\begin{tikzpicture}
		\begin{scope}[scale=0.4]
			\diskshadded{2}{2}{5}{2}{black}{white,opacity=0.4};
			\diskshadded{1}{1}{5}{1}{black}{white,opacity=0.4};
			\diskshadded{0}{0}{5}{0}{black}{white,opacity=0.4};
			\diskshadded{-1}{-1}{5}{-1}{black}{white,opacity=0.4};
			\diskshadded{-1}{-2}{4}{-2}{black}{white,opacity=0.4};
			\diskshadded{-1}{-3}{3}{-3}{black}{white,opacity=0.4};
			\diskshadded{-1}{-4}{2}{-4}{black}{white,opacity=0.4};
		
			\diskshadded{2}{1}{4}{1}{black}{gray,opacity=0.3};
			\diskshadded{1}{0}{4}{0}{black}{gray,opacity=0.3};
			\diskshadded{0}{-1}{4}{-1}{black}{gray,opacity=0.3};
			\diskshadded{0}{-2}{3}{-2}{black}{gray,opacity=0.3};
			\diskshadded{0}{-3}{2}{-3}{black}{gray,opacity=0.3};
			
			\wordbox{red}{2}{-1}{$v$};
		\end{scope}
	\begin{scope}[scale=0.4,xshift=8.5cm]
		\diskshadded{2}{2}{5}{2}{black}{white,opacity=0.4};
		\diskshadded{1}{1}{5}{1}{black}{white,opacity=0.4};
		\diskshadded{0}{0}{5}{0}{black}{white,opacity=0.4};
		\diskshadded{-1}{-1}{5}{-1}{black}{white,opacity=0.4};
		\diskshadded{-1}{-2}{4}{-2}{black}{white,opacity=0.4};
		\diskshadded{-1}{-3}{3}{-3}{black}{white,opacity=0.4};
		\diskshadded{-1}{-4}{2}{-4}{black}{white,opacity=0.4};
		\diskshadded{2}{-1}{4}{-1}{black}{gray,opacity=0.3};
		\diskshadded{3}{-2}{3}{-2}{black}{gray,opacity=0.3};
		\wordbox{red}{2}{-1}{$v$};
	\end{scope}
	\end{tikzpicture}
	\caption{A bulk site $v\in \mathfrak{D}_{\text{int}}$ and sites nearby. (Left) For the terms $\widetilde{K}^{\mathfrak{D}}_v$ and $\widetilde{K}^{\mathfrak{D}}_u$ to overlap, $u$ must be within the shaded region. (Right) The number of cases are reduced to a smaller number.}\label{fig:bulk_sites}
\end{figure}
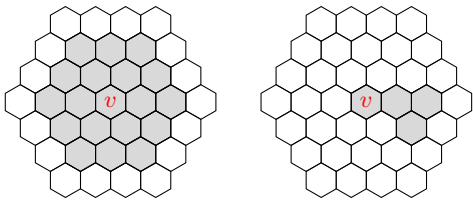

We show that the modular current vanishes in the interior of the disk $\mathfrak{D}$, \emph{i.e.,}
\begin{equation}\label{eq:bulk_current_appendix}
	f^{\mathfrak{D}}_{vu}=0,\quad \forall \,v, u\in \mathfrak{D}_{\text{int}}.
\end{equation}
 If $v$ and $u$ are sufficiently far apart so that the support of the operators $\widetilde{K}^{\mathfrak{D}}_v$ and $\widetilde{K}^{\mathfrak{D}}_u$ are non-overlapping,  $f^{\mathfrak{D}}_{vu}=0$ follows trivially. This reduces the number of nontrivial cases to be considered to a finite number; see Fig.~\ref{fig:bulk_sites}. 

The Markov decomposition \eqref{eq:markov_decomposition} and our coarse-graining procedure yield certain ``symmetries" of the problem that help us reduce the number of cases.\footnote{We remark that this ``symmetry" is due to the way that the coarse-graining is done as well as the area law assumption. The physical system does not need to possess any symmetry.} For example, the modular current are exactly the same under 6-fold rotations of the region, and thus it is enough to calculate $f_{vu}^\mathfrak{D}$ for $u$ to be four choices; see the right side of Fig.~\ref{fig:bulk_sites}. The calculations for these four choices of $u$ are further simplified thanks to the sign flip under reflection of the region.

For instance, to calculate $f_{vw}^{\mathfrak{D}}$, where ${v,w}$ is an edge of $\mathfrak{D}_{\text{int}}$ and $w$ is on the right of $v$.
\begin{equation}
	\tikz[scale=0.5,baseline=0ex,yshift=-0.15cm]{
		\diskshadded{1}{1}{1}{1}{black}{white,opacity=0.4};
		\diskshadded{2}{1}{2}{1}{black}{white,opacity=0.4};
		\draw[dotted, blue, line width=0.8 pt] (-0.5, 0.866) -- (2 +0.5, 0.866);
		\draw[line width=0.5pt, fill=black] (0.5, 0.866) circle (0.2);
		\draw[line width=0.5pt] (1.5, 0.866) circle (0.2);
		\node[] at (0.5, -0.05) {$v$};
		\node[] at (1.5, -0.05) {$w$};
	}
\end{equation}
 Here the dashed line passing through $v$ and $w$ is the axis of the reflection that is relevant. We shall need to calculate a linear combination of terms, including the following configurations:
\begin{equation}
	\begin{aligned}
			i\left\langle \left[
		\tikz[scale=0.25,baseline=0ex,yshift=-0.15cm]{
			\diskshadded{1}{1}{1}{1}{black}{white,opacity=0.4};
			\diskshadded{2}{1}{2}{1}{black}{white,opacity=0.4};
			\diskshadded{1}{0}{1}{0}{black}{white,opacity=0.4};
			\coloredke{yellow!50!red!50!white,opacity=0.5}{2}{1}{0};
			
			\draw[line width=0.5pt, fill=black] (0.5, 0.866) circle (0.2);
			\draw[line width=0.5pt] (1.5, 0.866) circle (0.2);
		}
		\,\, , \,
		\tikz[scale=0.25,baseline=0ex,yshift=-0.15cm]{
			\diskshadded{1}{1}{1}{1}{black}{white,opacity=0.4};
			\diskshadded{2}{1}{2}{1}{black}{white,opacity=0.4};
			\diskshadded{1}{0}{1}{0}{black}{white,opacity=0.4}; 
			\coloredke{yellow!50!red!50!white,opacity=0.5}{1}{1}{0};
			
			\draw[line width=0.5pt, fill=black] (0.5, 0.866) circle (0.2);
			\draw[line width=0.5pt] (1.5, 0.866) circle (0.2);
		} \right]\right\rangle
		&=J\\
		i\left\langle \left[
		\tikz[scale=0.25,baseline=0ex,yshift=-0.15cm]{
			%	\diskshadded{1}{1}{1}{1}{black}{white,opacity=0.4};
			\diskshadded{2}{1}{2}{1}{black}{white,opacity=0.4};
			\diskshadded{1}{0}{1}{0}{black}{white,opacity=0.4};
			\diskshadded{2}{0}{2}{0}{black}{white,opacity=0.4};
			\coloredke{yellow!50!red!50!white,opacity=0.5}{-2}{2}{1};
			
			\draw[line width=0.5pt, fill=black] (1,0 ) circle (0.2);
			\draw[line width=0.5pt] (2, 0) circle (0.2);
		}
		\,\, , \,
		\tikz[scale=0.25,baseline=0ex,yshift=-0.15cm]{
			%\diskshadded{1}{1}{1}{1}{black}{white,opacity=0.4};
			\diskshadded{2}{1}{2}{1}{black}{white,opacity=0.4};
			\diskshadded{1}{0}{1}{0}{black}{white,opacity=0.4}; 
			\diskshadded{2}{0}{2}{0}{black}{white,opacity=0.4};
			\coloredke{yellow!50!red!50!white,opacity=0.5}{-1}{2}{1};
			
			\draw[line width=0.5pt, fill=black] (1,0 ) circle (0.2);
			\draw[line width=0.5pt] (2, 0) circle (0.2);
		} \right]\right\rangle
		&=-J,
	\end{aligned}
\end{equation}
and
\begin{equation}
	\begin{aligned}
		i\left\langle \left[
	\tikz[scale=0.25,baseline=0ex,yshift=-0.15cm]{
		\diskshadded{1}{1}{1}{1}{black}{white,opacity=0.4};
		\diskshadded{2}{1}{2}{1}{black}{white,opacity=0.4};
		\diskshadded{1}{0}{1}{0}{black}{white,opacity=0.4};
		\diskshadded{2}{0}{2}{0}{black}{white,opacity=0.4};
		\coloredke{yellow!50!red!50!white,opacity=0.5}{-1}{1}{1};
		
		\draw[line width=0.5pt, fill=black] (1,0 ) circle (0.2);
		\draw[line width=0.5pt] (2, 0) circle (0.2);
	}
	\,\, , \,
	\tikz[scale=0.25,baseline=0ex,yshift=-0.15cm]{
		\diskshadded{1}{1}{1}{1}{black}{white,opacity=0.4};
		\diskshadded{2}{1}{2}{1}{black}{white,opacity=0.4};
		\diskshadded{1}{0}{1}{0}{black}{white,opacity=0.4}; 
		\diskshadded{2}{0}{2}{0}{black}{white,opacity=0.4};
		\coloredkf{green!50!yellow!50!white,opacity=0.5}{-2}{2}{1};
		
		\draw[line width=0.5pt, fill=black] (1,0 ) circle (0.2);
		\draw[line width=0.5pt] (2, 0) circle (0.2);
	} \right]\right\rangle
	&=J\\
	i\left\langle \left[
	\tikz[scale=0.25,baseline=0ex,yshift=-0.15cm]{
		\diskshadded{1}{1}{2}{1}{black}{white,opacity=0.4};
		\diskshadded{0}{0}{1}{0}{black}{white,opacity=0.4};
		\coloredke{yellow!50!red!50!white,opacity=0.5}{-2}{1}{1};
		
		\draw[line width=0.5pt, fill=black] (0.5, 0.866) circle (0.2);
		\draw[line width=0.5pt] (1.5, 0.866) circle (0.2);
	}
	\,\, , \,
	\tikz[scale=0.25,baseline=0ex,yshift=-0.15cm]{
		\diskshadded{1}{1}{2}{1}{black}{white,opacity=0.4};
		\diskshadded{0}{0}{1}{0}{black}{white,opacity=0.4};
		\coloredkf{green!50!yellow!50!white,opacity=0.5}{-3}{2}{1};
		
		\draw[line width=0.5pt, fill=black] (0.5, 0.866) circle (0.2);
		\draw[line width=0.5pt] (1.5, 0.866) circle (0.2);
	} \right]\right\rangle
	&=-J.
\end{aligned}
\end{equation}
Note that such terms come in pairs and the ``mirror image" gives the opposite contribution. Terms whose ``mirror image" is itself must be zero, e.g.:
\begin{equation}
		i\left\langle \left[\,
	\tikz[scale=0.3,baseline=0ex,yshift=0.32cm]{	
		\diskshadded{0}{0}{2}{0}{black}{white,opacity=0.4};
		\coloredke{yellow!50!red!50!white,opacity=0.5}{3}{1}{0};
		
		\draw[line width=0.5pt, fill=black] (0, 0) circle (0.2);
		\draw[line width=0.5pt] (1, 0) circle (0.2);
	}
	\,\, , \,
	\tikz[scale=0.3,baseline=0ex,yshift=0.32cm]{
		\diskshadded{0}{0}{2}{0}{black}{white,opacity=0.4};
		\coloredke{yellow!50!red!50!white,opacity=0.5}{0}{1}{0};
		
		\draw[line width=0.5pt, fill=black] (0, 0) circle (0.2);
		\draw[line width=0.5pt] (1, 0) circle (0.2);
	}\, \right]\right\rangle
	=0.
\end{equation}
Thus, the end result is zero, for sites in the interior of a disk. We arrive at Eq.~(\ref{eq:bulk_current_appendix}).

\subsubsection{Modular current in the vicinity of $\mathfrak{D}_{\partial}$}
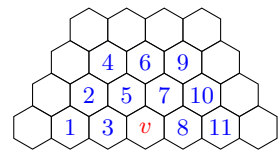
\begin{figure}[h]
		\centering	
		\begin{tikzpicture}[scale=0.5,baseline=0ex,yshift=-0.15cm]
			\diskshadded{2}{2}{5}{2}{black}{white,opacity=0.4};
			\diskshadded{1}{1}{5}{1}{black}{white,opacity=0.4};
			\diskshadded{0}{0}{5}{0}{black}{white,opacity=0.4};
			\diskshadded{-1}{-1}{5}{-1}{black}{white,opacity=0.4};
			
			\wordbox{red}{2}{-1}{$v$};
			%1
			\wordbox{blue}{0}{-1}{$1$};
			\wordbox{blue}{1}{-1}{$3$};
			\wordbox{blue}{3}{-1}{$8$};
			\wordbox{blue}{4}{-1}{$11$};
			%2
			\wordbox{blue}{1}{0}{$2$};
			\wordbox{blue}{2}{0}{$5$};
			\wordbox{blue}{3}{0}{$7$};
			\wordbox{blue}{4}{0}{$10$};
			%3
			\wordbox{blue}{2}{1}{$4$};
			\wordbox{blue}{3}{1}{$6$};
			\wordbox{blue}{4}{1}{$9$};
		\end{tikzpicture}
	\caption{A boundary site $v\in \mathfrak{D}_{\partial}$ and sites nearby.}\label{fig:boundary_sites}
\end{figure}

Below we calculate the modular current in the vicinity of the edge of the disk $\mathfrak{D}$. The boundary breaks the ``reflection symmetry" of the coarse-grained lattice, and the modular current may obtain a nonzero value. The calculation strategy is similar to the above. Due to locality, the modular current can only flow from $v$ to nearby sites; see Fig.~\ref{fig:boundary_sites}, where the nearby sites are numbered.

For the choice of $u$ and $v$ in the Fig.~\ref{fig:boundary_sites}, we calculated all the relevant values. The numbers are summarized in Table~\ref{table:bdy_values}. 
\begin{table}[h!]
	\centering
	\begin{tabular}{|| c || c | c | c | c | c | c | c | c | c | c | c ||} 
		\hline
		$u$ &1&2&3&4&5&6&7&8&9&10&11 \\
		\hline
		$f_{vu}^{\mathfrak{D}}$ &$\frac{-J}{9}$&$\frac{-J}{18}$&$\frac{5J}{36}$&$\,\,0\,\,$&
		$\,\,0\,\,$&$\,\,0\,\,$&$\,\,0\,\,$&$\frac{-5J}{36}$
		&$\,\,0\,\,$&$\frac{J}{18}$&$\frac{J}{9}$ \\
		 [0.5 ex] 
		\hline
	\end{tabular}
	\caption{The modular current involving a boundary site $v$.}
	\label{table:bdy_values}
\end{table}

Because all these calculations are similar, we present a simple but nontrivial example, i.e., the calculation that leads to  $f_{v1}^{\mathfrak{D}}=-J/9$. 
%Equivalently, we show $f_{1v}=J/9$.
\begin{equation}
	\begin{aligned}
		f_{1v}^{\mathfrak{D}}
		&= i\left\langle \left[-\frac{
		\tikz[scale=0.22,baseline=0ex,yshift=-0.15cm]{
			\diskshadded{1}{1}{2}{1}{black}{white,opacity=0.4};
			\diskshadded{0}{0}{2}{0}{black}{white,opacity=0.4};
			\coloredke{yellow!50!red!50!white,opacity=0.5}{-3}{1}{0};
		}
	}{2}
		\,\, , \,
		\frac{\tikz[scale=0.22,baseline=0ex,yshift=-0.15cm]{
			\diskshadded{1}{1}{2}{1}{black}{white,opacity=0.4};
			\diskshadded{0}{0}{2}{0}{black}{white,opacity=0.4};
			\coloredkf{green!50!yellow!50!white,opacity=0.5}{0}{1}{0};
		} 
	}{3}
	\right]\right\rangle
	+i\left\langle \left[\frac{\tikz[scale=0.22,baseline=0ex,yshift=-0.15cm]{
			\diskshadded{1}{1}{2}{1}{black}{white,opacity=0.4};
			\diskshadded{0}{0}{2}{0}{black}{white,opacity=0.4};
			\coloredkf{green!50!yellow!50!white,opacity=0.5}{0}{0}{0};
		} 
	}{3}		
	\,\, , \,
	-\frac{\tikz[scale=0.22,baseline=0ex,yshift=-0.15cm]{
	\diskshadded{1}{1}{2}{1}{black}{white,opacity=0.4};
	\diskshadded{0}{0}{2}{0}{black}{white,opacity=0.4};
	\coloredke{yellow!50!red!50!white,opacity=0.5}{0}{1}{0};
}
}{2}
	\right]\right\rangle\\
&	
+i\left\langle \left[\,
-\frac{\tikz[scale=0.22,baseline=0ex,yshift=0.32cm]{	
	\diskshadded{0}{0}{2}{0}{black}{white,opacity=0.4};
	\coloredke{yellow!50!red!50!white,opacity=0.5}{3}{1}{0};
}
}{2}
\,\, , \,
-\frac{\tikz[scale=0.22,baseline=0ex,yshift=0.32cm]{
	\diskshadded{0}{0}{2}{0}{black}{white,opacity=0.4};
	\coloredke{yellow!50!red!50!white,opacity=0.5}{0}{1}{0};
}
}{2}
\, \right]\right\rangle
+i\left\langle \left[
\frac{\tikz[scale=0.22,baseline=0ex,yshift=-0.15cm]{
		\diskshadded{1}{1}{2}{1}{black}{white,opacity=0.4};
		\diskshadded{0}{0}{2}{0}{black}{white,opacity=0.4};
		\coloredkf{green!50!yellow!50!white,opacity=0.5}{-2}{1}{1};
	}
}{3}
\,\, , \,
\frac{\tikz[scale=0.24,baseline=0ex,yshift=-0.15cm]{
		\diskshadded{1}{1}{2}{1}{black}{white,opacity=0.4};
		\diskshadded{0}{0}{2}{0}{black}{white,opacity=0.4};
		\coloredkf{green!50!yellow!50!white,opacity=0.5}{0}{1}{0};
	}
}{3} \right]\right\rangle\\
&=  \frac{-1}{6}\times 0 + \frac{-1}{6}\times 0 + \frac{1}{4}\times 0 +\frac{1}{9}\times J	\\
&= \frac{J}{9}.
	\end{aligned}
\end{equation}
To obtain the right side of the first equation, we have used the fact that commutators of non-overlapping terms as well as terms of the form $i\langle [K_{XY}, K_Y]\rangle$ are zero. On the right side of the second equation, the zeros comes from the vanishing of conditional mutual information ($I(X:Z\vert Y)_{\rho}=0$ implies $J(X,Y,Z)=0$); the $J$ in the last term is obtained using the topological invariance, which deforms the regions to the $A,B,C$ partition of a disk (Fig.~\ref{fig:abc} in the main text). All the other terms in Table~\ref{table:bdy_values} can be  obtained in a similar manner.

\end{document}